\documentclass[pra,twocolumn,tightenlines,showpacs,nofootinbib]{revtex4}
\usepackage{bm,dcolumn,amsmath,graphicx}

\newcommand{\eref}[1]{Eq.~(\ref{#1})}
\newcommand{\tref}[1]{Table~\ref{#1}}

\begin{document}

\title{Transition frequency shifts with fine-structure
constant variation for {Yb~II}}

\author{S. G. Porsev$^{1,2}$}
\author{V. V. Flambaum$^1$}
\author{J. R. Torgerson$^3$}
\affiliation{$^1$ School of Physics, University of New South Wales,
                  Sydney, NSW 2052, Australia}
\affiliation{$^2$ Petersburg Nuclear Physics Institute, Gatchina,
                  Leningrad district, 188300, Russia}
\affiliation{$^3$ University of California, Los Alamos National Laboratory,
Physics Division, P.O. Box 1663, Los Alamos, New Mexico 87545, USA}

\date{ \today }
\pacs{31.15.A-, 06.20.Jr, 31.15.am}

\begin{abstract}
In this paper we report calculations of the relativistic corrections to
transition frequencies ($q$ factors) of Yb~\textsc{ii} for the
transitions from the odd-parity states to the metastable state
$4f^{13}6s^2\, ^2\!F_{7/2}^o$. These transitions are of particular
interest experimentally since they possess some of the largest $q$
factors calculated to date and the $^2\!F_{7/2}^o$ state can be prepared
with high efficiency. This makes Yb~\textsc{ii} a very attractive
candidate for the laboratory search for variation of the fine-structure
constant $\alpha$.
\end{abstract}

\maketitle

\section{Introduction}
\label{sec_intro}

A discovery of acceleration of the universe (see,
e.g.,~\cite{CopSamTsu06}) is usually associated with the existence of
the dark energy. The latter, according to the theories describing
cosmological evolution, may be a reason of variations of the fundamental
constants. There is an on-going discussion in the literature whether the
fine-structure constant $\alpha$ could change during evolution of the
universe or not. The Australian group reported in~\cite{MurWebFla03} a
nonzero result while other astrophysical groups do not confirm
it~\cite{QuaReiLev04,SriChaPet04}. But as it was argued
in~\cite{MurWebFla07}, a more thorough analysis of the data used in
~\cite{SriChaPet04} also leads to a nonzero result. New laboratory and
astrophysical investigations are in progress.

Laboratory studies of hypothetical variation of the fine-structure
constant are based on the fact that transition frequencies in
atoms depend on $\alpha Z$, where $Z$ is the atomic number. Supposing
that the modern value of $\alpha$ differs from its value in the
earlier universe we can find relativistic transition frequencies shifts,
determined by so-called $q$ factors, according to
\begin{align}
\omega = \omega_\mathrm{lab} + q x, \quad x \equiv
\left({\alpha}/{\alpha_\mathrm{lab}}\right)^2 - 1\,.
\label{qfactor1}
\end{align}
Most advantageous for these studies are the atoms and ions for which $q$
factors of transitions between certain states significantly differ from
each other.

In~\cite{DzuFlaWeb99}, it was proposed to use transitions whose $q$
factors are large and of opposite sign for laboratory measurements. In
particular, it was shown that a good choice for an experiment would be
Hg$^+$ because the $q$ factor of the transition from the ground state to
the low-lying $^2\!D_{5/2}$ state is very large and negative. There are
two main considerations for the choice of a second transition for a
comparison: 1) it should be convenient for an experiment; 2) its $q$
factor should be small or positive because the measured quantity is
proportional to the difference of $q/\omega$ of the transitions being
compared.

At present the best laboratory constraint on the temporal variation of
$\alpha$ of $\dot{\alpha}/\alpha = (-1.6 \pm 2.3) \times 10^{-17}/$year
was obtained by Rosenband {\it et al.} in Ref.~\cite{RosHumSch08} by
comparing the frequencies of the $^1\!S_0 \rightarrow \, ^3\!P^o_0$
transition in $^{27}$Al$^+$ and the $^2\!S_{1/2} \rightarrow \,
^2\!D_{5/2}$ transition in $^{199}$Hg$^+$.

In this paper we propose to use Yb~\textsc{ii} for future experiments
because there are several transitions whose wavelengths can be
synthesized with modern laser technologies and whose $q$ factors are
very large and of opposite signs.

Yb~\textsc{ii} has two low-lying metastable states $4f^{14} 5d\,
^2\!D_{5/2}$ (24333\,cm$^{-1}$) and $4f^{13}6s^2\, ^2\!F_{7/2}^o$ (21419
cm$^{-1}$) decaying to the ground state $4f^{14} 6s\, ^2\!S_{1/2}$
through electric quadrupole and electric octupole transitions,
correspondingly. The $q$ factors of the $^2\!D_{5/2}$  and
$^2\!F_{7/2}^o$ states in respect to the ground state $^2\!S_{1/2}$ were
calculated in~\cite{DzuFlaWeb99,DzuFlaMar03,DzuFla08Yb}. For the
former $q \approx 10000$\,cm$^{-1}$ and for the latter
$q \approx -60000$\,cm$^{-1}$. In this work we carry out calculation of the $q$ factors for
certain experimentally interesting odd-parity states with respect to the
metastable state $4f^{13}6s^2\, ^2\!F_{7/2}^o$. Combining the results
obtained in this work with the results presented
in~\cite{DzuFlaWeb99,DzuFlaMar03,DzuFla08Yb}, we can easily find $q$
factors ranging from $-60000$\,cm$^{-1}$ to $+ 75000$\,cm$^{-1}$
associated with experimentally accessible transitions. This observation
can potentially increase the sensitivity of the measurements for
Yb~\textsc{ii} to $\alpha$ variation by more than a factor of two
compared to all previously considered comparisons and makes this ion a
very good candidate to establish a record constraint on $\alpha$ variaton
in laboratory studies.

The paper is organized as follows. Sec.~\ref{sec_method} is devoted to
the method of calculation of the properties of Yb~\textsc{ii}. In
Sec.~\ref{results} we discuss results and the experimental possibilities.
Sec.~\ref{conclusions} contains concluding remarks. Atomic
units ($\hbar = |e| = m_e = 1$) are used throughout the paper.

\section{Method of calculation}
\label{sec_method}

To find $q$ factors we need to solve the atomic relativistic eigenvalue
problem for different values of $\alpha$ or, respectively, for different
values of $x$ from~\eref{qfactor1}. The value of $x$ can be chosen
somewhat arbitrarily, but two conditions should be satisfied. It should
be sufficiently small to neglect nonlinear corrections and sufficiently
large to make calculations numerically stable. Our experience shows that
the choice of $|x|=1/8$ allows us to meet both conditions.

Thus, we have to calculate atomic frequencies $\omega_\pm$ for two
values $x=\pm 1/8$ of the parameter $x$. The corresponding $q$ factor is
given by
\begin{align}
q = 4 (\omega_+ - \omega_-) .
\label{qfactor2}
\end{align}

The ground state configuration of Yb~\textsc{ii} is ($1s^2 ...\, 4f^{14}
6s$). The configuration of the first excited state of Yb~\textsc{ii} is
($1s^2 ...\, 4f^{13} 6s^2$). This is a metastable state because there is
only weak ($E3$) transition from this state to the ground state. For
experimental purposes this metastable state lives sufficiently long to
be treated as a ``ground'' state. Since we are interested in calculation
of $q$ factors with respect to the metastable state with the open $f$
shell, we have to treat Yb~\textsc{ii} as an ion with 15 electrons above
closed shells. This makes calculations of Yb~\textsc{ii} rather
complicated.

In this paper we have carried out pure {\it ab initio} calculations in
the frame of the fifteen-electron CI method. The [$1s^2 ...\, 5p^6$]
electrons are treated as core electrons while $4f$, $6s$, and $5d$
electrons are in the valence space.

We started by solving the Dirac-Hartree-Fock (DHF) equations. The
self-consistency procedure was done for ($1s^2 ...\, 4f^{13} 6s^2$)
configuration. After that, the $5d_{3/2}$ and $5d_{5/2}$ orbitals were
constructed as follows: all electrons were frozen and one electron from
the $6s$ shell was moved to the $5d$ shell, thus constructing the
valence orbitals $5d_{3/2}$ and $5d_{5/2}$ for the $4f^{13}\, 5d6s$
configuration.

In the next stage, we constructed virtual orbitals using the method
described in~\cite{Bog91,KozPorFla96} and applied by us for calculating
different properties of Fe~\textsc{i} and
Fe~\textsc{ii}~\cite{PorKosTup07,PorKozRei09}. In this method, an upper
component of virtual orbitals is formed from the previous orbital of the
same symmetry by multiplication by some smooth function of radial
variable $r$. The lower component is then formed using the kinetic
balance condition.

Our basis sets included $s,p,d,$ and $f$ orbitals with principle quantum
number $n\le N_{1,2}$ which we designate as $[N_1sp\,N_2df]$. We carried
out the calculations of energy levels, $g$, and $q$ factors in a
one-configurational (DHF) approximation and for two basis sets with
($N_1=7$ and $N_2=5$) and ($N_1=8$ and $N_2=6$). Configuration space was
formed with single and double excitations from the configurations
$4f^{13} 6s^2$, $4f^{13} 5d6s$, and $4f^{13} 5d^2$. All results which we
discuss below are obtained in the pure Coulomb approximation (i.e, the
Breit interaction was not included).

\section{Results and discussion}
\label{results}
The results of the 15-electron CI calculations of the frequencies in
respect to the metastable state $4f^{13} 6s^2\, ^2\!F^o_{7/2}$ were obtained in the
DHF approximation and for the basis sets $[7sp5df]$ and $[8sp6df]$.
These results are given in~\tref{Tab_w}.

As is seen from the table, the values found for the $[7sp5df]$ basis set
are in a good agreement with the experimental frequencies. For low-lying
states this agreement is at the level of a few percent. For higher-lying
levels the agreement is slightly worse but does not exceed 10\%. The
reason is that the configuration interaction is more significant for the
high-lying states, as can be seen by comparison of the results
obtained for the $[7sp5df]$ and $[8sp6df]$ basis sets. For the
$[8sp6df]$ basis set an agreement between theoretical and experimental
values for the high-lying states belonging to the configuration
$4f^{13}\,5d^2$ becomes much better. Note that the
configuration space constructed for the $[8sp6df]$ basis set included
$\sim 2\,000\,000$ determinants and the calculations
of energies were rather lengthy and time-consuming for this basis so a further increase
of the configuration space is impractical.
\begin{table}[bt]
\caption{The fifteen-electron CI calculations of the transition frequencies
$\omega$ (in\,cm$^{-1}$) for different basis sets. The Dirac-Coulomb Hamiltonian in the
frozen-core approximation is used. The transition frequencies
are calculated with respect to the metastable $4f^{13}6s^2\,\, ^2\!F^o_{7/2}$ state.
The experimental data from~\cite{NIST} are given in the last row.}

\label{Tab_w}

\begin{ruledtabular}

\begin{tabular}{lcccccc}
Config.& Term & $J$
&\multicolumn{1}{c}{DHF} &\multicolumn{1}{c}{$[7sp5df]$}
&\multicolumn{1}{c}{$[8sp6df]$} &\multicolumn{1}{c}{Exp.}

\\[1mm]
\hline\\
$4f^{13}(^2\!F^o) 6s^2$       &    $^2F^o$   &  7/2  &    0  &   0   &   0   &   0  \\[0.5mm]
$4f^{13}(^2\!F^o_{7/2}) 5d6s$ & $^3[11/2]^o$ &  9/2  & 5701  & 8648  & 7761  & 8805 \\
                              & $^3[11/2]^o$ & 11/2  & 5810  & 8835  & 7985  & 9144 \\
                              & $^3[11/2]^o$ & 13/2  & 6148  & 9359  & 8615  & 10213\\
                              &  $^3[5/2]^o$ &  7/2  & 7493  & 10564 & 9831  & 10561\\
                              &  $^3[7/2]^o$ &  9/2  & 8977  & 12086 & 11312 & 11633\\
                              &  $^3[9/2]^o$ &  7/2  & 9603  & 12578 & 11737 & 12076\\
                              & $^1[11/2]^o$ & 11/2  & 15143 & 13791 & 12686 & 13366\\
                              &  $^3[9/2]^o$ &  9/2  & 10613 & 13713 & 12940 & 13600\\
                              &  $^3[7/2]^o$ &  7/2  & 10531 & 13641 & 12901 & 13640\\
                              &  $^3[9/2]^o$ & 11/2  & 10834 & 14645 & 13575 & 14413\\
                              &  $^1[7/2]^o$ &  7/2  & 18688 & 17166 & 15786 & 16098\\
$4f^{13}(^2\!F^o_{7/2}) 5d6s$ &  $^1[9/2]^o$ &  7/2  & 19610 & 17782 & 16384 & 16923\\[0.5mm]
$4f^{13}(^2\!F^o_{5/2}) 5d6s$ &  $^3[9/2]^o$ &  7/2  & 15713 & 18904 & 18031 & 18617\\
                              &  $^3[9/2]^o$ &  9/2  & 16599 & 19734 & 18866 & 19499\\
                              &  $^3[9/2]^o$ & 11/2  & 17936 & 21171 & 20414 & 21496\\
$4f^{13}(^2\!F^o_{5/2}) 5d6s$ &  $^3[5/2]^o$ &  7/2  & 21092 & 24366 & 23634 & 23019\\
$4f^{13} 5d^2$                &  $^o$        &  7/2  & 22889 & 25847 & 23507 & 23854\\
$4f^{13}(^2\!F^o_{5/2}) 5d6s$ &  $^1[9/2]^o$ &  9/2  & 26117 & 25510 & 23633 & 23916\\
$4f^{13}(^2\!F^o_{5/2}) 5d6s$ &  $^3[7/2]^o$ &  7/2  & 20831 & 23799 & 22972 & 24011\\
$4f^{13}(^2\!F^o_{5/2}) 5d6s$ &  $^3[7/2]^o$ &  9/2  & 21626 & 24719 & 24412 & 24751\\
$4f^{13}(^2\!F^o_{7/2}) 5d^2$ &  $^3[7/2]^o$ &  9/2  & 23391 & 26320 & 24005 & 24936\\
$4f^{13}(^2\!F^o_{7/2}) 5d^2$ & $^3[13/2]^o$ & 11/2  & 24585 & 27315 & 24679 & 25129
\end{tabular}
\end{ruledtabular}
\end{table}

At the same time, as it follows from Tables~\ref{Tab_g} and \ref{Tab_q} where
the results for $g$ and $q$ factors are presented, both these quantities are rather
insensitive to the size of the configuration space. Using the experimental $g$ factors
and comparing them with the calculated ones, we are able to identify properly the
calculated energy levels. For the majority of the calculated levels
there is a very satisfactory agreement between theoretical and experimental
$g$ factors. This means that the configuration interaction is represented in
a proper way for these states and also means that the $q$ factors obtained for them
are correct.

Comparing the values of $g$ and $q$ factors obtained for the $[7sp5df]$
and $[8sp6df]$ basis sets, we see that almost all of them agree to each other at
the level of few per cent. An exception is the levels with $J=11/2$ belonging to the
terms $^1[11/2]^o$ and  $^3[9/2]^o$. These two  states are closer in the $[7sp5df]$
basis set approximation than in real experimental data so as to result in an artificial
mixing of these levels and, consequently, to a change in their $g$ and
$q$ factors.

\begin{table}[bt]
\caption{The fifteen-electron CI calculations of $g$ factors
for different basis sets. The Dirac-Coulomb Hamiltonian in the
frozen-core approximation is used. The experimental data presented
in the last row are taken from~\cite{NIST}.}

\label{Tab_g}

\begin{ruledtabular}

\begin{tabular}{lcccccl}
Config.& Term & $J$
&\multicolumn{1}{c}{DHF} &\multicolumn{1}{c}{$[7sp5df]$}
&\multicolumn{1}{c}{$[8sp6df]$} &\multicolumn{1}{c}{Exp.}
\\[1mm]
\hline\\
$4f^{13}(^2\!F^o) 6s^2$       &    $^2F^o$   &  7/2  & 1.143  & 1.443 & 1.143 & 1.145 \\[0.5mm]
$4f^{13}(^2\!F^o_{7/2}) 5d6s$ & $^3[11/2]^o$ &  9/2  & 0.934  & 0.933 & 0.933 & 0.935 \\
                              & $^3[11/2]^o$ & 11/2  & 1.122  & 1.121 & 1.121 & 1.112 \\
                              & $^3[11/2]^o$ & 13/2  & 1.231  & 1.231 & 1.231 & 1.230 \\
                              &  $^3[5/2]^o$ &  7/2  & 1.377  & 1.375 & 1.371 & 1.331 \\
                              &  $^3[7/2]^o$ &  9/2  & 1.296  & 1.286 & 1.278 & 1.264 \\
                              &  $^3[9/2]^o$ &  7/2  & 1.027  & 1.010 & 0.995 & 0.991 \\
                              & $^1[11/2]^o$ & 11/2  & 1.093  & 1.195 & 1.131 & 1.119 \\
                              &  $^3[9/2]^o$ &  9/2  & 1.132  & 1.140 & 1.147 & 1.158 \\
                              &  $^3[7/2]^o$ &  7/2  & 1.048  & 1.066 & 1.084 & 1.124 \\
                              &  $^3[9/2]^o$ & 11/2  & 1.244  & 1.143 & 1.206 & 1.214 \\
                              &  $^1[7/2]^o$ &  7/2  & 1.130  & 1.120 & 1.121 & 1.119 \\
$4f^{13}(^2\!F^o_{7/2}) 5d6s$ &  $^1[9/2]^o$ &  9/2  & 1.079  & 1.088 & 1.089 & 1.093 \\[0.5mm]
$4f^{13}(^2\!F^o_{5/2}) 5d6s$ &  $^3[9/2]^o$ &  7/2  & 0.720  & 0.720 & 0.718 & 0.720 \\
                              &  $^3[9/2]^o$ &  9/2  & 0.984  & 0.983 & 0.982 & 0.967 \\
                              &  $^3[9/2]^o$ & 11/2  & 1.127  & 1.128 & 1.129 & 1.115 \\
$4f^{13}(^2\!F^o_{5/2}) 5d6s$ &  $^3[5/2]^o$ &  7/2  & 1.002  & 1.020 & 1.045 & 1.10  \\
$4f^{13} 5d^2$                &  $^o$        &  7/2  & 1.295  & 1.297 & 1.277 & 1.18  \\
$4f^{13}(^2\!F^o_{5/2}) 5d6s$ &  $^1[9/2]^o$ &  9/2  & 0.945  & 0.993 & 1.000 & 1.01  \\
$4f^{13}(^2\!F^o_{5/2}) 5d6s$ &  $^3[7/2]^o$ &  7/2  & 1.168  & 1.158 & 1.144 & 1.150 \\
$4f^{13}(^2\!F^o_{5/2}) 5d6s$ &  $^3[7/2]^o$ &  9/2  & 1.147  & 1.092 & 1.086 & 1.10  \\
$4f^{13}(^2\!F^o_{7/2}) 5d^2$ &  $^3[7/2]^o$ &  9/2  & 1.311  & 1.310 & 1.309 & 1.29  \\
$4f^{13}(^2\!F^o_{7/2}) 5d^2$ & $^3[13/2]^o$ & 11/2  & 0.957  & 0.955 & 0.955 & 0.97
\end{tabular}
\end{ruledtabular}
\end{table}
As is seen from \tref{Tab_q} the levels belonging to the $4f^{13} 5d^2$ configuration
have the largest $q$ factors. This is not surprising because this configuration differs
by two electrons from the configuration of the metastable state $4f^{13} 6s^2$. When
the fine structure constant $\alpha$ tends to its nonrelativistic limit, one-electron
energies of the $6s$ and $5d$ electrons change in different directions, which leads to
an increase of the $q$ factors. Note in this respect that the high-lying levels which
nominally belong to
the $4f^{13} 6s5d$ configuration have (in reality) a rather large admixture of the
$4f^{13} 5d^2$ configuration. As a consequence their $q$ factors are also large.

We can estimate the accuracy of the calculated $q$ factors as a
difference between the largest and smallest values (in each line) listed in \tref{Tab_q} for
three basis sets. Such a conservative estimate shows that the accuracy of our
calculations is not worse than 10\%.
\begin{table}[bt]
\caption{The fifteen-electron CI calculations of the $q$ factors (in\,cm$^{-1}$)
for different basis sets. The Dirac-Coulomb Hamiltonian in the
frozen-core approximation is used.
The $q$ factors are calculated with respect to the metastable
$4f^{13}6s^2\,\, ^2\!F^o_{7/2}$ state.}

\label{Tab_q}

\begin{ruledtabular}

\begin{tabular}{lccccc}
Config.& Term & $J$
&\multicolumn{1}{c}{DHF} &\multicolumn{1}{c}{$[7sp5df]$}
&\multicolumn{1}{c}{$[8sp6df]$}
\\[1mm]
\hline\\
$4f^{13}(^2\!F^o_{7/2}) 5d6s$ & $^3[11/2]^o$ &  9/2  & 14522  &  14458 &  14118 \\
                              & $^3[11/2]^o$ & 11/2  & 14843  &  14813 &  14496 \\
                              & $^3[11/2]^o$ & 13/2  & 15396  &  15548 &  15371 \\
                              &  $^3[5/2]^o$ &  7/2  & 14563  &  14481 &  14167 \\
                              &  $^3[7/2]^o$ &  9/2  & 14736  &  14690 &  14402 \\
                              &  $^3[9/2]^o$ &  7/2  & 14876  &  14761 &  14469 \\
                              & $^1[11/2]^o$ & 11/2  & 17113  &  15910 &  16573 \\
                              &  $^3[9/2]^o$ &  9/2  & 15372  &  15462 &  15306 \\
                              &  $^3[7/2]^o$ &  7/2  & 15518  &  15568 &  15396 \\
                              &  $^3[9/2]^o$ & 11/2  & 15270  &  16988 &  16015 \\
                              &  $^1[7/2]^o$ &  7/2  & 17892  &  17698 &  17691 \\
$4f^{13}(^2\!F^o_{7/2}) 5d6s$ &  $^1[9/2]^o$ &  9/2  & 17243  &  17541 &  17577 \\[0.5mm]
$4f^{13}(^2\!F^o_{5/2}) 5d6s$ &  $^3[9/2]^o$ &  7/2  & 23242  &  23388 &  23072 \\
                              &  $^3[9/2]^o$ &  9/2  & 23795  &  23968 &  23650 \\
                              &  $^3[9/2]^o$ & 11/2  & 24610  &  24951 &  24767 \\
$4f^{13}(^2\!F^o_{5/2}) 5d6s$ &  $^3[5/2]^o$ &  7/2  & 24243  &  24568 &  24439 \\
$4f^{13} 5d^2$                &  $^o$        &  7/2  & 27078  &  27201 &  26910 \\
$4f^{13}(^2\!F^o_{5/2}) 5d6s$ &  $^1[9/2]^o$ &  9/2  & 26185  &  26562 &  25918 \\
$4f^{13}(^2\!F^o_{5/2}) 5d6s$ &  $^3[7/2]^o$ &  7/2  & 24611  &  24180 &  23976 \\
$4f^{13}(^2\!F^o_{5/2}) 5d6s$ &  $^3[7/2]^o$ &  9/2  & 24432  &  25180 &  25521 \\
$4f^{13}(^2\!F^o_{7/2}) 5d^2$ &  $^3[7/2]^o$ &  9/2  & 27489  &  27711 &  27638 \\
$4f^{13}(^2\!F^o_{7/2}) 5d^2$ & $^3[13/2]^o$ & 11/2  & 26804  &  26816 &  26607
\end{tabular}
\end{ruledtabular}
\end{table}
In \tref{Tab_q_final} we present the recommended values of the $q$ factors found
in this work. For future reference, we also list in the table the $q$ factors
of the lowest-lying even-parity $4f^{14}\, 5d\,\, ^2D_{3/2,5/2}$ states  and
the  $4f^{13}\, 6s^2\,\, ^2F^o_{7/2}$ state found in Refs.~\cite{DzuFlaWeb99,DzuFlaMar03,DzuFla08Yb}.
These values were obtained with respect to the ground state. Using the $q$ factors
found in this work with respect to the metastable state and $q(^2\!F^o_{7/2})$ obtained
in~\cite{DzuFlaMar03,DzuFla08Yb} with respect to the ground state, it easy to recalculate
the $q$ factors presented in \tref{Tab_q} with respect to the ground state. It can be done
just by subtracting $q(^2\!F^o_{7/2})$ from these $q$ factors because
for positive $\omega = \omega_1 - \omega_2$ the value of $q$ is equal to $(q_1 - q_2)$.

Experimentally one can search for a variation of $\alpha$ by comparing
two frequencies of atomic transitions over a long period of time.
Following the Ref.~\cite{DzuFlaMar03} we can represent a
measured quantity $\Delta(t)$ as
\begin{equation}
\Delta(t) =  \frac{d}{dt} \left( {\rm ln} \frac{\omega_1}{\omega_2} \right)
          =   \frac{\dot{\omega_1}}{\omega_1} -
              \frac{\dot{\omega_2}}{\omega_2}  ,
\label{Delt}
\end{equation}
where $\dot{\omega} \equiv d\omega/dt$.
Taking into account \eref{qfactor1} we can rewrite \eref{Delt} as follows
\begin{equation}
\Delta(t) \approx \left( \frac{2 q_1}{\omega_1} - \frac{2 q_2}{\omega_2} \right)
             \left( \frac{\dot{\alpha}}{\alpha_{\rm lab}} \right) .
\label{Del_t}
\end{equation}
Of particular interest experimentally are narrow transitions with large
$q$ values. Many of these exist from the $^2\!F^o_{7/2}$ metastable
state to higher-lying states with wavelengths which can be synthesized
with modern laser technologies. As is seen in \tref{Tab_q_final}, a
very large $\Delta(t)$ can be expected if the frequencies of the
$4f^{14} 5d\,\, ^2\!D_{3/2}\, (22961\, {\rm cm}^{-1}) -
4f^{13} 6s^2\,\, ^2\!F^o_{7/2}\, (21419\, {\rm cm}^{-1})$  and
$4f^{14} 6s\,\, ^2\!S_{1/2}$ (the ground state)
$-\, 4f^{13} 6s^2\,\, ^2\!F^o_{7/2}\, (21419\, {\rm cm}^{-1})$ transitions are compared.
The $q$ factor of the former ($q_1$) is positive
($q_1 \approx 71000\, {\rm cm}^{-1}$) and the $q$ factor of the latter ($q_2$)
is negative ($q_2 \approx -60000\, {\rm cm}^{-1}$).
The transition $4f^{13} 6s^2\,\, ^2\!F^o_{7/2} -
4f^{14} 5d\,\, ^2\!D_{5/2}$ has nearly the same $q$ value as the
$4f^{13} 6s^2\,\, ^2\!F^o_{7/2} - 4f^{14} 5d\,\, ^2\!D_{3/2}$ transition
and the necessary wavelength at 3.4\,$\mu$m is somewhat easier to
synthesize than the 6.5\,$\mu$m to the $^2\!D_{3/2}$ state. Another
consideration is that the transition to $^2\!D_{3/2}$ at 6.5 $\mu$m is
forbidden for even isotopes, but is weakly allowed through the hyperfine
interaction within odd isotopes. In either case, transitions from
$m_F=0\rightarrow m_F'=0$ in odd isotopes are preferred due to the
smaller (and quadratic) Zeeman shift to the transitions relative to the
linear shift within the even isotopes.

Substituting the $q$ values into \eref{Del_t} for the two most sensitive
transitions described above, and using the estimate
$|\dot{\alpha}/\alpha_{\rm lab}| < 10^{-16} \,\, {\rm yr}^{-1}$
\cite{RosHumSch08} we find
\begin{equation}
\Delta(t) < 10^{-14} \,\,{\rm yr}^{-1} .
\end{equation}
It is worth noting that a presence of transitions for which the $q$
factors are very large and have opposite sign is favorable. First, it
leads to increasing $\Delta(t)$ (as it follows from  \eref{Del_t}) and,
second, allows better control of some systematic errors that are not
correlated with signs and magnitudes of the frequency shifts.

It is also important to see that transitions from the
$4f^{13} 6s^2\,\, ^2\!F^o_{7/2}$ state to higher-lying
states with $J>7/2$ possess more easily synthesized wavelengths.
States with $J>7/2$ with energies less than 48000\,cm$^{-1}$ are most
likely all metastable since there are no even-parity states with
$J>5/2$ below this energy. The $q$-factors are somewhat lower, but this
consideration may be outweighed by practical considerations such as
available sources of narrowband laser light. The value of $\Delta(t)$
for a comparison between $4f^{14}\, 6s\,\, ^2S_{1/2} - 4f^{13}\,
6s^2\,\, ^2F^o_{7/2}$ and one of these transitions is still
significantly greater than any other comparison considered to date.

Preparation of the metastable state $4f^{13}6s^2\, ^2\!F^o_{7/2}$ has
already been performed in the laboratory by several groups including
one of us (JRT). One of the
simplest schemes to populate this state is from the $^2\!D_{3/2}$ state
at 22961\,cm$^{-1}$ which is populated by spontaneous decay from
$^2\!P^o_{1/2}$ at 27062\,cm$^{-1}$; the upper state of the most
accessible laser cooling transition. Transitions from $^2\!D_{3/2}$ at
861 nm, 1062 nm and 1163 nm  can all be used with varying efficiencies.
For example, we have used a simple external cavity diode laser at 861 nm
to prove the technique. For any of these wavelengths, state preparation
is limited by the 5 ms lifetime of the $^2\!D_{5/2}$ state at 24333
cm$^{-1}$.

\begin{table*}[bt]
\caption{The recommended values of the $q$-factors (in\,cm$^{-1}$)
found in this work with respect to the $4f^{13} 6s^2\,\,^2\!F^o_{7/2}$ state
and the $q$ factors found in Refs.~\cite{DzuFla08Yb,DzuFlaMar03,DzuFlaWeb99} with respect
to the ground state $4f^{14} 6s\,\, ^2\!S_{1/2}$.}

\label{Tab_q_final}

\begin{ruledtabular}

\begin{tabular}{lcccccc}
Config.& Term & $J$
&\multicolumn{1}{c}{This work} &\multicolumn{1}{c}{Ref.~\cite{DzuFla08Yb}}
&\multicolumn{1}{c}{Ref.~\cite{DzuFlaMar03}} &\multicolumn{1}{c}{Ref.~\cite{DzuFlaWeb99}}
\\[1mm]
\hline\\
$4f^{14} 5d$                  &    $^2D$     &  3/2  &        &        &  10118 & 12582  \\
                              &    $^2D$     &  5/2  &        &        &  10397 & 11438  \\
$4f^{13}(^2\!F^o) 6s^2$       &    $^2F^o$   &  7/2  &        & -63752 & -56737 &        \\
\hline \\
$4f^{13}(^2\!F^o_{7/2}) 5d6s$ & $^3[11/2]^o$ &  9/2  & 14100  &        &        &         \\
                              & $^3[11/2]^o$ & 11/2  & 14500  &        &        &         \\
                              & $^3[11/2]^o$ & 13/2  & 15400  &        &        &         \\
                              &  $^3[5/2]^o$ &  7/2  & 14200  &        &        &         \\
                              &  $^3[7/2]^o$ &  9/2  & 14400  &        &        &         \\
                              &  $^3[9/2]^o$ &  7/2  & 14500  &        &        &         \\
                              & $^1[11/2]^o$ & 11/2  & 16600  &        &        &         \\
                              &  $^3[9/2]^o$ &  9/2  & 15300  &        &        &         \\
                              &  $^3[7/2]^o$ &  7/2  & 15400  &        &        &         \\
                              &  $^3[9/2]^o$ & 11/2  & 16000  &        &        &         \\
                              &  $^1[7/2]^o$ &  7/2  & 17700  &        &        &         \\
$4f^{13}(^2\!F^o_{7/2}) 5d6s$ &  $^1[9/2]^o$ &  9/2  & 17600  &        &        &         \\[0.5mm]
$4f^{13}(^2\!F^o_{5/2}) 5d6s$ &  $^3[9/2]^o$ &  7/2  & 23100  &        &        &         \\
                              &  $^3[9/2]^o$ &  9/2  & 23700  &        &        &         \\
                              &  $^3[9/2]^o$ & 11/2  & 24800  &        &        &         \\
$4f^{13}(^2\!F^o_{5/2}) 5d6s$ &  $^3[5/2]^o$ &  7/2  & 24400  &        &        &         \\
$4f^{13} 5d^2$                &  $^o$        &  7/2  & 26900  &        &        &         \\
$4f^{13}(^2\!F^o_{5/2}) 5d6s$ &  $^1[9/2]^o$ &  9/2  & 25900  &        &        &         \\
$4f^{13}(^2\!F^o_{5/2}) 5d6s$ &  $^3[7/2]^o$ &  7/2  & 24000  &        &        &         \\
$4f^{13}(^2\!F^o_{5/2}) 5d6s$ &  $^3[7/2]^o$ &  9/2  & 25500  &        &        &         \\
$4f^{13}(^2\!F^o_{7/2}) 5d^2$ &  $^3[7/2]^o$ &  9/2  & 27600  &        &        &         \\
$4f^{13}(^2\!F^o_{7/2}) 5d^2$ & $^3[13/2]^o$ & 11/2  & 26600  &        &        &
\end{tabular}
\end{ruledtabular}
\end{table*}

\section{Conclusion}
\label{conclusions}

We have calculated relativistic frequency shifts ($q$ factors) for a
number of transitions from excited states of Yb~\textsc{ii} to the
metastable state $4f^{13}6s^2\,\, ^2\!F^o_{7/2}$ state. We found that
all these $q$ factors are large ($\sim 10^4$\,cm$^{-1}$) and positive.
The $q$ factor for the transition $^2\!S_{1/2} -\, ^2\!F_{7/2}^o$
is very large and negative. Because the
$^2\!F_{7/2}^o$ state is convenient for an experiment and can be
prepared with high efficiency, we conclude that Yb~\textsc{ii} is a very
good candidate for the laboratory search for possible variation of the
fine-structure constant $\alpha$.

\section{acknowledgments}
This work was supported by Australian Research Council and Marsden grant.
The work of SGP was supported in part by the Russian Foundation for Basic
Research under Grants No. 07-02-00210-a and No. 08-02-00460-a and the work
of JRT was supported at LANL by a Laboratory Directed Research and Development grant.


\begin{thebibliography}{14}
\expandafter\ifx\csname natexlab\endcsname\relax\def\natexlab#1{#1}\fi
\expandafter\ifx\csname bibnamefont\endcsname\relax
  \def\bibnamefont#1{#1}\fi
\expandafter\ifx\csname bibfnamefont\endcsname\relax
  \def\bibfnamefont#1{#1}\fi
\expandafter\ifx\csname citenamefont\endcsname\relax
  \def\citenamefont#1{#1}\fi
\expandafter\ifx\csname url\endcsname\relax
  \def\url#1{\texttt{#1}}\fi
\expandafter\ifx\csname urlprefix\endcsname\relax\def\urlprefix{URL }\fi
\providecommand{\bibinfo}[2]{#2}
\providecommand{\eprint}[2][]{\url{#2}}

\bibitem[{\citenamefont{Copeland et~al.}(2006)\citenamefont{Copeland, Sami, and
  Tsujikawa}}]{CopSamTsu06}
\bibinfo{author}{\bibfnamefont{E.~J.} \bibnamefont{Copeland}},
  \bibinfo{author}{\bibfnamefont{M.}~\bibnamefont{Sami}}, \bibnamefont{and}
  \bibinfo{author}{\bibfnamefont{S.}~\bibnamefont{Tsujikawa}},
  \bibinfo{journal}{Int. J. Mod. Phys. D} \textbf{\bibinfo{volume}{15}},
  \bibinfo{pages}{1753} (\bibinfo{year}{2006}).

\bibitem[{\citenamefont{Murphy et~al.}(2003)\citenamefont{Murphy, Webb, and
  Flambaum}}]{MurWebFla03}
\bibinfo{author}{\bibfnamefont{M.~T.} \bibnamefont{Murphy}},
  \bibinfo{author}{\bibfnamefont{J.~K.} \bibnamefont{Webb}}, \bibnamefont{and}
  \bibinfo{author}{\bibfnamefont{V.~V.} \bibnamefont{Flambaum}},
  \bibinfo{journal}{Mon. Not. R. Astron. Soc.} \textbf{\bibinfo{volume}{345}},
  \bibinfo{pages}{609} (\bibinfo{year}{2003}).

\bibitem[{\citenamefont{Quast et~al.}(2004)\citenamefont{Quast, Reimers, and
  Levshakov}}]{QuaReiLev04}
\bibinfo{author}{\bibfnamefont{R.}~\bibnamefont{Quast}},
  \bibinfo{author}{\bibfnamefont{D.}~\bibnamefont{Reimers}}, \bibnamefont{and}
  \bibinfo{author}{\bibfnamefont{S.~A.} \bibnamefont{Levshakov}},
  \bibinfo{journal}{Astron. \& Astrophys.} \textbf{\bibinfo{volume}{415}},
  \bibinfo{pages}{L7} (\bibinfo{year}{2004}).

\bibitem[{\citenamefont{Srianand et~al.}(2004)\citenamefont{Srianand, Chand,
  Petitjean, and Aracil}}]{SriChaPet04}
\bibinfo{author}{\bibfnamefont{R.}~\bibnamefont{Srianand}},
  \bibinfo{author}{\bibfnamefont{H.}~\bibnamefont{Chand}},
  \bibinfo{author}{\bibfnamefont{P.}~\bibnamefont{Petitjean}},
  \bibnamefont{and} \bibinfo{author}{\bibfnamefont{B.}~\bibnamefont{Aracil}},
  \bibinfo{journal}{Phys. Rev. Lett.} \textbf{\bibinfo{volume}{92}},
  \bibinfo{pages}{121302} (\bibinfo{year}{2004}).

\bibitem[{Mur()}]{MurWebFla07}
\bibinfo{note}{M. T. Murphy, J. K. Webb, and V. V. Flambaum, Phys. Rev. Lett.
  {\bf 99}, 239001 (2007); Mon. Not. R. Astron. Soc. {\bf 345}, 609 (2007)}.

\bibitem[{\citenamefont{Dzuba et~al.}(1999)\citenamefont{Dzuba, Flambaum, and
  Webb}}]{DzuFlaWeb99}
\bibinfo{author}{\bibfnamefont{V.~A.} \bibnamefont{Dzuba}},
  \bibinfo{author}{\bibfnamefont{V.~V.} \bibnamefont{Flambaum}},
  \bibnamefont{and} \bibinfo{author}{\bibfnamefont{J.~K.} \bibnamefont{Webb}},
  \bibinfo{journal}{Phys. Rev. A} \textbf{\bibinfo{volume}{59}},
  \bibinfo{pages}{230} (\bibinfo{year}{1999}).

\bibitem[{\citenamefont{Rosenband et~al.}(2008)\citenamefont{Rosenband, Hume,
  Schmidt, Chou, Brusch, Lorini, Oskay, Drullinger, Fortier, Stalnaker
  et~al.}}]{RosHumSch08}
\bibinfo{author}{\bibfnamefont{T.}~\bibnamefont{Rosenband}},
  \bibinfo{author}{\bibfnamefont{D.~B.} \bibnamefont{Hume}},
  \bibinfo{author}{\bibfnamefont{P.~O.} \bibnamefont{Schmidt}},
  \bibinfo{author}{\bibfnamefont{C.~W.} \bibnamefont{Chou}},
  \bibinfo{author}{\bibfnamefont{A.}~\bibnamefont{Brusch}},
  \bibinfo{author}{\bibfnamefont{L.}~\bibnamefont{Lorini}},
  \bibinfo{author}{\bibfnamefont{W.~H.} \bibnamefont{Oskay}},
  \bibinfo{author}{\bibfnamefont{R.~E.} \bibnamefont{Drullinger}},
  \bibinfo{author}{\bibfnamefont{T.~M.} \bibnamefont{Fortier}},
  \bibinfo{author}{\bibfnamefont{J.~E.} \bibnamefont{Stalnaker}},
  \bibnamefont{et~al.}, \bibinfo{journal}{Science}
  \textbf{\bibinfo{volume}{319}}, \bibinfo{pages}{1808} (\bibinfo{year}{2008}).

\bibitem[{\citenamefont{Dzuba et~al.}(2003)\citenamefont{Dzuba, Flambaum, and
  Marchenko}}]{DzuFlaMar03}
\bibinfo{author}{\bibfnamefont{V.~A.} \bibnamefont{Dzuba}},
  \bibinfo{author}{\bibfnamefont{V.~V.} \bibnamefont{Flambaum}},
  \bibnamefont{and} \bibinfo{author}{\bibfnamefont{M.~V.}
  \bibnamefont{Marchenko}}, \bibinfo{journal}{Phys. Rev. A}
  \textbf{\bibinfo{volume}{68}}, \bibinfo{pages}{022506}
  (\bibinfo{year}{2003}).

\bibitem[{\citenamefont{Dzuba and Flambaum}(2008)}]{DzuFla08Yb}
\bibinfo{author}{\bibfnamefont{V.~A.} \bibnamefont{Dzuba}} \bibnamefont{and}
  \bibinfo{author}{\bibfnamefont{V.~V.} \bibnamefont{Flambaum}},
  \bibinfo{journal}{Phys. Rev. A} \textbf{\bibinfo{volume}{77}},
  \bibinfo{pages}{012515} (\bibinfo{year}{2008}).

\bibitem[{\citenamefont{Bogdanovich}(1991)}]{Bog91}
\bibinfo{author}{\bibfnamefont{P.}~\bibnamefont{Bogdanovich}},
  \bibinfo{journal}{Lith. Phys. J.} \textbf{\bibinfo{volume}{31}},
  \bibinfo{pages}{79} (\bibinfo{year}{1991}).

\bibitem[{\citenamefont{Kozlov et~al.}(1996)\citenamefont{Kozlov, Porsev, and
  Flambaum}}]{KozPorFla96}
\bibinfo{author}{\bibfnamefont{M.~G.} \bibnamefont{Kozlov}},
  \bibinfo{author}{\bibfnamefont{S.~G.} \bibnamefont{Porsev}},
  \bibnamefont{and} \bibinfo{author}{\bibfnamefont{V.~V.}
  \bibnamefont{Flambaum}}, \bibinfo{journal}{J. \ Phys. \ B}
  \textbf{\bibinfo{volume}{29}}, \bibinfo{pages}{689} (\bibinfo{year}{1996}).

\bibitem[{\citenamefont{Porsev et~al.}(2007)\citenamefont{Porsev, Koshelev,
  Tupitsyn, Kozlov, Reimers, and Levshakov}}]{PorKosTup07}
\bibinfo{author}{\bibfnamefont{S.~G.} \bibnamefont{Porsev}},
  \bibinfo{author}{\bibfnamefont{K.~V.} \bibnamefont{Koshelev}},
  \bibinfo{author}{\bibfnamefont{I.~I.} \bibnamefont{Tupitsyn}},
  \bibinfo{author}{\bibfnamefont{M.~G.} \bibnamefont{Kozlov}},
  \bibinfo{author}{\bibfnamefont{D.}~\bibnamefont{Reimers}}, \bibnamefont{and}
  \bibinfo{author}{\bibfnamefont{S.~A.} \bibnamefont{Levshakov}},
  \bibinfo{journal}{Phys. Rev. A} \textbf{\bibinfo{volume}{76}},
  \bibinfo{pages}{052507} (\bibinfo{year}{2007}).

\bibitem[{\citenamefont{Porsev et~al.}(2009)\citenamefont{Porsev, Kozlov, and
  Reimers}}]{PorKozRei09}
\bibinfo{author}{\bibfnamefont{S.~G.} \bibnamefont{Porsev}},
  \bibinfo{author}{\bibfnamefont{M.~G.} \bibnamefont{Kozlov}},
  \bibnamefont{and} \bibinfo{author}{\bibfnamefont{D.}~\bibnamefont{Reimers}},
  \bibinfo{journal}{Phys. Rev. A} \textbf{\bibinfo{volume}{79}},
  \bibinfo{pages}{032519} (\bibinfo{year}{2009}).

\bibitem[{NIS()}]{NIST}
\emph{\bibinfo{title}{{\rm NIST,} {A}tomic {S}pectra {D}atabase}},
  \urlprefix\url{http://physics.nist.gov/cgi-bin/AtData/main_asd}.

\end{thebibliography}


\end{document}